\documentstyle[12pt,epsfig,amsfonts]{article}

\makeatletter

\@addtoreset{equation}{section} \makeatother

\newcommand{\nn}{\nonumber}
\newcommand{\be}{\begin{equation}}
\newcommand{\ee}{\end{equation}}
\newcommand{\ba}{\begin{eqnarray}}
\newcommand{\ea}{\end{eqnarray}}
\newcommand{\dle}[1]{\label{#1}}
\newcommand{\dla}[1]{\label{#1}}
\newcommand{\dr}[1]{\ref{#1}}
\newcommand{\dc}[1]{\cite{#1}}

\newcommand{\dnote}[1]{}

\newcommand{\gsim}{\raise.3ex\hbox{$>$\kern-.75em\lower1ex\hbox{$\sim$}}}
\newcommand{\lsim}{\raise.3ex\hbox{$<$\kern-.75em\lower1ex\hbox{$\sim$}}}

\newcommand{\veca}{{\mathbf{a}}}
\newcommand{\vecb}{{\mathbf{b}}}
\newcommand{\vecx}{{\mathbf{x}}}
\newcommand{\vecy}{{\mathbf{y}}}
\newcommand{\vecz}{{\mathbf{z}}}
\newcommand{\paa}{\partial}
\newcommand{\sqrtg}{\sqrt{-\gamma}}
\newcommand{\si}{{\sigma}}
\newcommand\spr[1]{\mathaccent19{#1}}
\renewcommand\={\!\!\!&=&\!\!\!}

\begin{document}

\renewcommand{\thefootnote}{\fnsymbol{footnote}}


\begin{center}

{\large\bf Self-intersections and gravitational properties of
chiral cosmic strings in Minkowski space\vskip 0.1cm}

\vskip 1.2cm {\large D.A.\ Steer\footnote{E-mail: {\tt
Daniele.Steer@physics.unige.ch}}}\\ \vskip 5pt \vskip 3pt
D\'epartement de Physique Th\'eorique, Universit\'e de Gen\`eve, \\
24 Quai Ernest Ansermet,
1211 Gen\`eve 4, Switzerland \\
\vskip 0.3cm
\end{center}

\vskip 1.2cm

\renewcommand{\thefootnote}{\arabic{footnote}}
\setcounter{footnote}{0} \typeout{--- Main Text Start ---}

\begin{abstract}

Chiral cosmic strings are naturally produced at the end of D-term
inflation and they may have interesting cosmological consequences.
As was first proved by Carter and Peter, the equations of motion
for chiral cosmic strings in Minkowski space are integrable (just
as for Nambu-Goto strings).  Their solutions are labeled by a
function $k(\sigma - t)$ where $t$ is time and $\sigma$ is the
invariant length along the string, and the constraints on $k$,
which determines the charge on the string, are that $0 \leq k^2
\leq 1$. We review the origin of this parameter and also
discuss some general properties of such strings which can be
deduced from the equations of motion.  The metric around infinite
chiral strings is then constructed in the weak field limit, and
studied as a function of $k$.  We also consider the angular
momentum of circular chiral loops, and extend previous work to
consider the evolution and self-intersection properties of a more
general family of chiral cosmic string loops for which
$k^2(\sigma-t)$ is not constant.

\end{abstract}

\section{Introduction}

In the last few years the scenario of structure formation from
cosmic strings has become increasingly tenuous, since its
predictions differ significantly from the new high accuracy measurements of
the temperature fluctuations in the cosmic
microwave background radiation.  Most studies of
such observational consequences of strings have focused on
structureless Nambu-Goto (NG) strings \dc{James1,CHM,avelino,Levon,CMS}
and global strings \dc{Neil,Ruth}, and in each case the recent predictions
are based on numerical simulations of the evolution of the string network
postulated to form at the GUT phase transition.
One should recall though that there are some unresolved and
potentially important uncertainties in the simulations --- it is
very difficult, for example, to resolve the very disparate scales
which characterize the the network, as well as to deal with
gravitational backreaction effect --- and hence a combination of
numerical work with analytical modeling \dc{James1,Levon,CMS} has
also been used to make predictions from NG strings.

Our focus here is not on NG strings but rather on
{\em chiral cosmic strings}.
These strings are a type of current carrying string \dc{Witten}
for which the world-sheet current $j^{i}$ is null;
$$
j^{i}j_i = j^2 = 0.
$$
(Here $i=(0,1)$ and the 2D world sheet metric $\gamma_{ij}$ defined
below raises and lowers indices.)
One motivation for studying such chiral strings comes from the well
known SUSY D-term inflation model.  In this model, strings are produced
at the end of inflation \dc{Rachel}
so that both mechanisms contribute to producing
density fluctuations.  However, the strings produced are chiral cosmic
strings and not NG strings \dc{DDT1}.  Hence in order to make
predictions for the $C_l$'s from this `strings plus inflation' model,
the evolution and cosmological consequences
of chiral cosmic string networks must be
understood. (There may exist models in which the strings formed at the
end of inflation are NG ones, however this is not true of D-term inflation.
In the case of `inflation plus NG strings', predictions may be found in
\dc{CHM2}.)

There are a number of differences between the properties of chiral
cosmic strings and NG strings.  One such regards the evolution of
the strings themselves: the null current on chiral strings can, as
in the case of other current-carrying strings, lead to the
formation of non self-intersecting stable loops called
vortons\footnote{As will be come clearer later, by a vorton we
mean a stable loop of arbitrary shape that never self-intersects.
This definition is different from that of Martins and Shellard
\dc{MS} who also require that these loops move
non-relativistically, suggesting that otherwise the charge on the
loops could be `thrown off'.  We are not able to comment on such a
mechanism, however see \dc{DDDD} for a discussion of the
scattering of zero-modes from chiral strings.}. This is
potentially catastrophic as the energy density in the chiral
string network could quickly dominate the energy density in the
universe if stable vortons are present.  It is therefore important
to see if vortons are produced, and in section \dr{s:kvar} we
study the self-intersection properties of a family of chiral
cosmic string loops.  Another difference between NG and chiral
strings is that these line-like sources of energy generate
different metrics about them (section \dr{s:metric})\footnote{I am
aware that this comment disagrees with one I made in \dc{Verb}!  I
would like to thank Patrick Peter and Tanmay Vachaspati for
pointing out an error in my previous determination of the
metric.}.  One might therefore expect them to produce different
perturbations in the matter and radiation through which they pass.

Recently a number of steps have been made which allow for a
quantitative study of chiral cosmic string dynamics. First, a well
defined unique 2D effective action exists for these
strings \dc{MS,CP}.  From this action it was shown, with suitable
gauge choices, that the equations of motion are integrable in
Minkowski space \dc{CP} (see also \dc{US,BP} for
different presentations of the same result).  They are
\be
\frac{\paa^2 \vecx}{\paa t^2} - \frac{\paa^2 \vecx}{\paa \sigma^2}
= 0 \qquad \Longrightarrow \qquad \vecx(t,\si) =
\frac{1}{2}[\veca(t+\si) + \vecb(t-\si)],
\dle{eqnlecon}
\ee
where $t$ is background time, and $\sigma$ measures the invariant
length or energy along the string as in the NG case \dc{US}. The
constraints are
\ba
\spr{\veca}^2 \= 1,
\dla{c1}
\\
\spr{\vecb}^2  \!\!\!& \le &\!\!\! 1,
\dla{c2}
\end{eqnarray}
where for instance $\spr\veca(q)\equiv d\veca(q)/dq.$  If one
defines
\be
k^2:= \spr{\vecb}^2
\dle{kddef}
\ee
so that $k=k(t-\sigma)$,
then it can be shown that $k^2$ determines the conserved
charge on the string (see also below).  Furthermore, if $k = $
constant $=1$ then this charge vanishes as required, since
$\spr{\vecb}=1$ is just the Nambu-Goto limit. In reference
\dc{US}, the self-intersection properties of chiral cosmic string
loops were also studied in the special case of $k =$
constant. In particular the strings were shown never to
self-intersect for $k=0$: this case corresponds to maximal charge
on the strings and to vorton solutions.

Here that work is extended, though we still consider Minkowski
space (with metric $\eta_{\mu \nu} = (+,-,-,-)$) throughout.
First, for completeness, we indicate in section \dr{s:2} how the
equations of motion (\dr{eqnlecon})-(\dr{c2}) are obtained from
the chiral action and how the charge mentioned above is defined.
This necessarily follows parts of reference \dc{US} rather
closely, though a small error in that paper is corrected.  We also
compare the chiral charge with the charges used for more general
current carrying strings.  In section \dr{s:angmmt} we summarize
some properties of chiral cosmic strings which result from the
equations of motion.  The metric around infinite chiral strings is
then studied as a function of $k$ and we comment in possible
consequences it may have for structure formation and CMB anisotropies
from chiral
cosmic strings. In section \dr{ss:ang}, the effect of angular
momentum on the motion of circular loops is considered by looking
at the effective potential introduced in \dc{CPG}. In section
\dr{s:kvar} we investigate the self-intersection properties of
loops with non-constant $k$. Finally conclusions are given in
section \dr{s:conc}.

\section{Review of chiral string equations of motion and
charges}\dle{s:2}

\subsection{Action and charges}\dle{s:action}

The effective 2D chiral string action has 2 terms: the
first is the usual NG action, and the second results from the zero
modes moving along the string.  Let $\phi$ be a dimensionless real
scalar field (the phase of the charge carriers) living on the 2D
string world sheet labeled by coordinates $\sigma^{i}$.
Then the action, which was first proposed by Carter and
Peter \dc{CP}, is
\begin{equation}
S = - \int d^2 \sigma\, \sqrt{-\gamma} \left(m^2 - \frac{1}{2}
\psi^2 \gamma^{ij} \phi_{,i} \phi_{,j}\right)
\dle{Baction}
\end{equation}
where $\gamma_{ij} = \eta_{\mu \nu}x^{\mu}_{,i} x^{\nu}_{,j}$ is
the induced world sheet metric and $x^{\mu}(\sigma^0,\sigma^1)$
the position of the string.  The dimensionless Lagrange multiplier
$\psi^2$ sets the constraint
\begin{equation}
\gamma^{ij} \phi_{,i} \phi_{,j} = 0 \qquad
\Longrightarrow \qquad   \frac{1}{\sqrtg}\paa_i(\sqrtg
\gamma^{ij}\phi_{,j})  = 0
\dle{wrong}
\end{equation}
so that $J^i = \gamma^{ij}\phi_{,j}$
is a conserved null current.  The
equation of motion $\delta S /
\delta \phi=0$ defines another conserved null current
$z^i$ by
\begin{equation}
\paa_i (\sqrtg \psi^2 \gamma^{ij} \phi_{,j})=0 \qquad
\Longrightarrow \qquad  z^i = \psi^2 \gamma^{ij}
\phi_{,j}.
\dle{currcon1}
\end{equation}
As noted in \dc{US}, the action (\dr{Baction}) in fact has an
infinite number of null conserved currents $j^i = f(\phi)
\phi^{,i}$ since equations (\dr{wrong}) and (\dr{currcon1}) imply
that $\psi = \psi(\phi)$.
The degeneracy of currents is broken by observing that
(\ref{Baction}) is invariant not only under coordinate
reparametrizations $\si^i\to\tilde\si^i=\tilde\si^i(\si^j)$ but
also under transformations
\begin{equation}
\phi\to\tilde\phi(\phi),\qquad{\rm with}\qquad
\psi\to\tilde\psi=\left({d\tilde\phi\over d\phi}\right)^{-1}\psi.
\dle{phitrans}
\end{equation}
These freedoms are removed making gauge choices (see \dc{CP,US} and
below), so that the only definition of current which is invariant under
(\dr{phitrans}) and hence independent of gauge choice is
\begin{equation}
 j^{i} = \psi \phi^{,i}.
\label{currT}
\end{equation}
This is null and conserved and, from Green's theorem, the
corresponding conserved charge is
\be
C = \int d\sigma^i \epsilon_{ik}j^{k},
\dle{Qfinal}
\ee
where $\epsilon$ is the antisymmetric surface measure tensor whose
square gives the induced metric; $\gamma_{ij} =
\epsilon_{ik}\epsilon^{k}_{ \; j}$ \dc{CP}.

For current-carrying strings with time-
or space-like currents, this degeneracy of possible conserved
charges is broken.  These strings are characterized by two
independent conserved quantum numbers (see for example \dc{CPG}).
The first, $Z$, is defined through the Noether current $z^i$ given
in (\dr{currcon1}):
$Z
=\int d\sigma^i \epsilon_{ik}z^{k}.$
The second, $N$ the integer winding number, is defined by
the topological current $\tilde{j}^{i} =
\frac{1}{2\pi} \epsilon^{ij}\phi_{,j}$ which is
automatically conserved in 1+1D:
$N = \int d\sigma^i
\epsilon_{ik}\tilde{j}^{k} = \frac{1}{2\pi}\int d\phi$ ($\phi$ is defined
modulo $2 \pi$).
As noted above, {\em neither} of these currents and
corresponding charges are gauge invariant for chiral strings.
The chiral charge $C$
is closely related to $N$ and $Z$ if one works in a gauge in
which $\psi(\phi)$ is constant:
on defining $\kappa_0 = \psi^2$ then
\be
C = \frac{Z}{\sqrt{\kappa_0}} = 2 \pi \sqrt{\kappa_0} N \qquad
(\psi(\phi) \; = \; {\rm constant}).
\ee
This gauge was in particular chosen in reference
\dc{MS}\footnote{Of course if $\psi =$ constant, $\phi$  can
always rescaled in the action such that $\psi^2 = 1/2\pi$ and $N=Z
= 2 \pi C$ as is usually assumed in the study of
vortons.}.\footnote{We have labeled the chiral charge by $C$ as
for circular loops it
coincides with the Bernoulli-type constant of motion considered in
\dc{CPG}.} As was discussed in detail in
\dc{CP,US}, and as we now summarize briefly, the equations of
motion resulting from (\dr{Baction}) simplify greatly in a gauge
for which $\psi(\phi)$ is not constant. (Indeed in this gauge,
$\psi(\phi)$ is closely related to function $k$ mentioned in the
introduction --- see below.)  Then there is no simple relation
between $N$, $Z$ and $C$, and one must work with this latter
gauge-independent charge.

\subsection{Equations of motion}\dle{s:eqn}

As was discussed in \dc{CP,US}, the equation of motion obtained by
varying the action with respect to $x^{\mu}$;
\begin{equation}
\paa_i \left[ \sqrtg \left( \gamma^{ij} + \frac{\psi^2}{m^2}
\phi^{,i} \phi^{,j} \right) x^{\mu}_{,j} \right]=0,
\dle{Mink}
\end{equation}
simplifies greatly if reparametrization invariance is used to
choose one of the coordinates to be $ \eta = m^{-1} \phi$.  It
then follows from (\dr{wrong}) that $\gamma^{\eta \eta} = 0$ and,
again as discussed in \dc{CP,US}, there is also freedom to choose
$\psi^2 = \gamma_{\eta \eta} = x_{,\eta} \cdot x_{,\eta}$. As a
result equation (\dr{Mink}) simplifies to
\be
\partial_q \partial_\eta x^{\mu} = 0,
\dle{eqnmot}
\ee
where the second world-sheet coordinate has been denoted by $q$.
Equation (\dr{eqnmot}) still allows the coordinates $q$ and $\eta$ each
to be transformed separately so that one can let
$$
q = t + \sigma \qquad , \qquad \eta = t - \sigma
$$
where $t=x^0$ is background time.  In that case the wave
equation (\dr{eqnmot}) takes the familiar form given in (\dr{eqnlecon}):
$$
\frac{\paa^2 \vecx}{\paa t^2} - \frac{\paa^2 \vecx}{\paa \sigma^2}
= 0 \qquad \Longrightarrow \qquad \vecx(t,\si) =
\frac{1}{2}[\veca(t+\si) + \vecb(t-\si)].
$$
The constraints coming from $\gamma^{\eta \eta} = 0$ and $\psi^2 =
\gamma_{\eta \eta}$ are respectively (\dr{c1}) and (\dr{c2}):
$$
\spr{\veca}^2(q)  = 1 \qquad , \qquad
 \spr{\vecb}^2(\eta) = k^2(\eta) \; \le \;
1.
$$
 Observe
that $\psi^2 =  x_{,\eta} \cdot x_{,\eta} =
(1-k^2(\eta))/4$.

In \dc{US} it was further shown that with these choices of coordinates, the stress energy tensor is given by
\be
T^{\mu \nu} (t,\vecy)= m^2 \int d\sigma \left( \dot x^{\mu} \dot
x^{\nu}  - x^{\mu}{}'x^{\nu}{}' \right) \delta^3(\vecy -
\vecx(t,\sigma)).
\dle{Tmunu}
\ee
Thus $E$, the constant energy, is given by $E = m^2 \int d\sigma$
so that $\sigma$ measures the energy or invariant length along the
string. Below, in section \dr{s:gen}, we will discuss the
contribution of the null current to the energy density, and the
metric around infinite chiral strings will also be considered
(section \dr{s:metric}).

Finally, in these ($t,\sigma$) coordinates, the charge $C$ is given by
\be
C = \int d\sigma \sqrt{-\gamma} j^{t} = \int d\sigma m
\psi(\sigma) = \frac{m}{2} \int d\sigma [1-k^2(\sigma)]^{1/2}
\dle{Cr}
\ee
and hence that it is determined by $k(\eta)$. The right hand side
of (\dr{Cr}) differs from the one given in \dc{US} by a factor of
2:  the reason is that $\sqrt{-\gamma}$ is coordinate dependent so
if $\gamma(\xi^0,\xi^1)$ denotes the determinant of the metric in
a specific $(\xi^0,\xi^1)$ coordinate system, then
$\sqrt{-\gamma(t,\sigma)} = 2 \sqrt{-\gamma(q,\eta)}$.  This
factor of 2 was missing in \dc{US}.\footnote{Equations (\dr{Cr})
and (\dr{Qfinal}) do indeed agree since in $(t,\sigma)$
coordinates, $\epsilon_{t \sigma} = \sqrt{-\gamma(t,\sigma)} = -
\epsilon_{\sigma t}$.}

\section{Properties of chiral strings, metrics, and angular
momentum}\dle{s:angmmt}

\subsection{Some general properties of chiral strings}\dle{s:gen}

As observed in \dc{US,BP}, it follows immediately from (\dr{eqnlecon})
that ${\vecx'} \neq 0$ and $|\dot{\vecx}| \neq 1$ so that there are
no cusps on chiral cosmic strings.

Also $\dot{\vecx} \neq 0$, though this does not mean that the
string cannot appear to be at rest, since the only visible
component of velocity is that perpendicular to the string. For
example, a static infinite chiral string parallel to the
$\hat{\vecz}$-axis is given by
$$
\veca = (t + \sigma) \hat{\vecz}, \qquad \vecb = -k (t - \sigma)
\hat{\vecz},
$$
where $k$ is constant.  These satisfy
(\dr{c2}) and give
\be
\vecx(t,\sigma)  = \frac{1}{2} [ t(1-k) + \sigma(1+k) ]
\hat{\vecz}.
\dle{gg}
\ee
In the NG limit ($k=1$), $\vecx = \sigma \hat{\vecz}$ so that points
of constant $\sigma$ are at fixed values of ${\bf \hat{z}}$ (and
$\dot{\vecx}=0$). For any $k < 1$ points of constant $\sigma$ move
along the $z$-axis with time and $\dot{\vecx} \neq 0$, though the
string itself never changes position.  Below, in section \dr{s:metric},
we will look
at $T^{\mu \nu}$ given in (\dr{Tmunu}) for the infinite string
(\dr{gg}) and hence consider the metric about the string.

In the particular case of the infinite string (\dr{gg}), $\dot{\vecx}$
and $\vecx'$ were parallel.  More generally, and again as noted in
\dc{US,BP}, for any arbitrary shaped cosmic string (infinite
or a loop), the limit $k=0$ $\forall \eta$ is special:  here
$\dot{\vecx} = \vecx'$ with $|\dot{\vecx}| = |\vecx'| = 1/2$.  Thus
the {\em only} component of velocity is parallel to the string which
moves along itself at half the speed of light.  The string, whatever its
shape, therefore
appears to be stationary and it can never self-intersect
\dc{US}.  If the string forms a loop, these are called vortons
(i.e.\ non-self-intersecting solutions which need not be
circular)\dnote{This vorton state has $k=0$ or $\psi = 1$ =
constant.  Hence in this case one can rescale $\phi$ such that
$N=Z$.} which radiate neither gravitational
energy nor gravitational angular momentum.

We note one minor difference between such `static' $k=0$ chiral
strings and static NG strings (which have $k=1$ and $\spr{\veca} = -
\spr{\vecb}$).  The physical length $\ell$ of the string is related to $\sigma$ by
$$ d\ell = \sqrt{-\gamma_{\sigma \sigma}} d\sigma = \frac{1}{2} \left[ - 1 +
k^2 + 2 (1 - \spr{\vecb} \cdot \spr{\veca} ) \right]^{1/2} d\sigma
$$
so that of
course $d \ell = d\sigma$ for static NG strings.  For static chiral
strings with $k=0$, $d\ell = d\sigma / 2$: the string energy is
equipartitioned between tension and angular momentum (due to the current)
as will be discussed in section \dr{ss:ang}.  From (\dr{Cr}) it follows that the charge $C$ on a vorton is given by
$$
C = m \int d\ell = m L_{\rm phys} \qquad \qquad (k=0),
$$
where $L_{\rm phys}$ is the constant physical length of the vorton.

\subsection{Metric around infinite chiral strings}\dle{s:metric}

From equation (\dr{Tmunu}), the stress energy tensor for the infinite string
given in (\dr{gg}) is
\be
T^{\mu \nu} = m^2 \left(
\begin{array}{cccc}  1 & 0 & 0 & \frac{1-k}{2} \\
0 & 0 & 0 & 0 \\
0 & 0 & 0 & 0 \\
\frac{1-k}{2} & 0 & 0 & -k
\end{array} \right) \delta(x) \delta(y) .
\dle{Tinf}
\ee
Note that $T^{00} \neq - T^{33}$ unless $k=1$ in which case the
off-diagonal terms also vanish.  These off-diagonal terms
represent the momentum along the string (in this case it is the
only momentum) given by $\dot{\vecx} = \frac{1}{2}(1-k)
\hat{\vecz}$.  For $k<1$, $T^{\mu \nu}$ cannot be put into
diagonal form by a Lorentz transformation along the string, as the
boost would have to be to a frame moving at the speed of light.  The
off-diagonal terms are a consequence of the null current on the
string.  (Off-diagonal terms are not present for space- or
time-like current carrying cosmic strings --- see, for example, \dc{PP}.)

Metrics generated by stress energy tensors of the form (\dr{Tinf})
have been considered in \dc{TM1,Tiglio,TM2}.  Here we comment on a few
properties of the weak-field metric obtained from (\dr{Tinf});
further details will be presented elsewhere \dc{StVa}.

In the weak field approximation, $g_{\mu \nu} = \eta_{\mu \nu} +
h_{\mu \nu}$ where $|h| \ll 1$, and in the de Donder gauge
$h_{\mu \nu}$ satisfies \dc{Wein}
\ba
\Box h_{\mu \nu} &=& 16 \pi G (T_{\mu \nu} - \frac{1}{2} \eta_{\mu \nu} T_{\alpha}^{\alpha} )
\nonumber
\\
& = & 8 \pi G m^2
 \left(
\begin{array}{cccc}  1-k & 0 & 0 & -(1-k) \\
0 & 1+k & 0 & 0 \\
0 & 0 & 1+k & 0 \\
-(1-k) & 0 & 0 & 1-k
\end{array} \right) \delta(x) \delta(y)
\dle{hh}.
\ea
On writing $r^2 = x^2 + y^2$, the solutions to (\dr{hh}) can be
written as
\ba
h_{tt} = - h_{tz} = h_{zz} =: X(r,k) = 4 G m^2 (1-k) \ln(r/r_0),
\dle{X}
\\
h_{xx} = h_{yy} =: Q(r,k) = 4 G m^2 (1+k) \ln(r/r_0),
\dle{Q}
\ea
where $r_0$ is an integration constant which can be thought of as
the width of the string.  The metric obtained from (\dr{X}) and
(\dr{Q}) can be simplified by using the familiar coordinate
transformation $(1-Q(r,k)) r^2 = (1-2Gm^2(1+k))^2 R^2 $
\dc{Vilenkin} which gives
\ba
ds^2 &=& dt^2 (1 + X(R,k)) - dz^2 (1 - X(R,k)) - dR^2 -
(1-2Gm^2(1+k))^2 R^2 d\theta^2
\nn
\\
& - & 2 X(R,k) dt \; dz.
\dle{string}
\ea
The first line of (\dr{string}) is familiar --- it is the metric
one obtains for wiggly NG cosmic strings which have $T^{00} \neq
-T^{33}$ but $T^{03} = 0$ \dc{HiKi}.  Just as in that case, the
coefficient of the $d\theta^2$ term gives a deficit angle
$$
\delta(k) = 4 \pi G m^2 (1+k)
$$
which is now $k$ dependent.  It is worth noting that $\delta$ is
maximal for non-charged NG strings ($k=1$) and takes its minimum
value when $k=0$ (vortons).

The equations of motion for non-relativistic particles (Newtonian
limit) in the metric (\dr{string}) can be straightforwardly
written down.  As expected, there is a Newtonian potential
$\Phi(R,k) = X(R,k)/2$ which leads to an attractive Newtonian
force
$$
F(R,k) = \frac{2 G (1-k)}{R}
$$
towards the string.  Again this force is $k$ dependent: it
vanishes for NG strings and is maximal when $k=0$.  Thus one might
expect chiral strings with a large charge to be more effective in
forming wakes than ones with a smaller charge \dc{StVa}.

The less familiar term in the metric (\dr{string}) is the last
one, $2 X(R,k) dt \; dz$.  (This vanishes both for wiggly and
straight NG strings.)  Whilst this term has no effect on the
motion of non-relativistic particles, it does affect the motion of
relativistic particles and in particular photons (see also
\dc{TM1,TM2}).  To see that, note from (\dr{string}) that
geodesics are characterised by three conserved quantities, the
energy $e$, angular momentum $L$ and $z$-component of momentum
$p_z$.  These are given respectively by
\ba
e & = & \dot{t}  (1 + X) - X \dot{z},
\nn
\\
L &=& (1-2Gm^2(1+k))^2 R^2 \dot{\theta},
\nn
\\
p_z &=& \dot{z}(1-X) +  X \dot{t},
\nn
\ea
where for simplicity we have written $X(R,k) = X$, and a dot means
derivative with respect to an affine parameter in the case of
photons, and proper-time for particles.
Consider now photons for which the equations of
motion are
\ba
\dot{t} &=& e(1-X) + p_z X,
\nn
\\
\dot{z} &=& p_z (1+X) - eX,
\nn
\\
\dot{\theta} &=& \frac{L}{(1-2Gm^2(1+k))^2 R^2},
\nn
\\
\dot{R}^2  &=& e^2 (1-X) - p_z^2 (1+X) + 2 e p_z X -
\frac{L^2}{R^2 (1-2Gm^2(1+k))^2}.
\nn
\ea
Combining $\dot{z}$ with $\dot{t}$ gives
\be
\frac{dz}{dt} = \frac{ p_z (1+X) - e X}{e(1-X) + p_z X}.
\dle{dzdt}
\ee
Suppose a photon travels in a plane perpendicular to the string at
some $R=R_0$ so that $dz/dt|_{R=R_0} = 0$, and denote $X(R_0,k) =
\tilde{X}$.  Substituting into (\dr{dzdt}) gives
$$
e =  \frac{p_z (1+\tilde{X})}{\tilde{X}}
$$
so that from (\dr{dzdt})
$$
\frac{dz}{dt} = \frac{ \tilde{X}- X}{1 - X +  \tilde{X}}.
$$
The denominator is positive and the numerator also for $R < R_0$.
Therefore as the photon moves towards the string it gets dragged
in the positive $z$ direction. (This effect vanishes in the NG
limit as then $X=\tilde{X} = 0$ from (\dr{X}).)

It would be interesting to understand the effect of this dragging
on the temperature anisotropy caused by a single chiral cosmic
string. In this weak field limit, a preliminary calculation seems
to suggest that there is no effect --- the only anisotropy is
caused by the deficit angle $\delta$ and is given by \dc{HiKi}
$$
\frac{\delta T}{T} = 4 \pi G m^2 (1+k) v \gamma,
$$
where $\gamma$ is the usual Lorentz factor, and $v$ is the
velocity of the string which moves perpendicular to the line
connecting the string and the source.  A complete calculation
would require one to go beyond the weak field approximation.  The effects on
the lensing produced by chiral strings could then also be
considered. This study is in progress \dc{StVa}.

\subsection{Angular momentum and loops}\dle{ss:ang}

In the rest of this paper we consider the dynamics of chiral
cosmic loops. First note that in this gauge, the fact that there
is a component of velocity along the string itself (since
$\dot{\vecx}\cdot \vecx'= (1-k^2(\eta))/4$) suggests that closed
strings --- loops --- will carry angular momentum. (Of course, NG
loops can also carry angular momentum). Recall next that a string
of invariant length $L$ forms a loop if
\be
\vecx(t,\si+L)=\vecx(t,\si).
\ee
In the centre of mass frame where $\oint d\sigma \; \dot{\vecx} =
0$ the functions $\veca$ and $\vecb$ are also periodic with period
$L$:  chiral strings like NG ones have periodic motion with
period $L/2$.  The vectors $\spr{\veca}$ and $\spr{\vecb}$ can be
expanded in a Fourier series; for $L=2 \pi$,
\ba
\spr{\veca}(q) & = &\sum_{n \geq 1} ({\bf A}_n \cos n q + {\bf
B}_n \sin n q ),
\nn
\\
\spr{\vecb}(\eta) & = &\sum_{n \geq 1} ({\bf C}_n \cos n \eta +
{\bf D}_n \sin n \eta ),
\dla{fourier}
\ea
and the constraints on ${\bf A}_n$ and ${\bf B}_n$ are such that
$\spr{\veca}^2=1$ (equation (\dr{c1})). The vectors ${\bf C}_n$
and ${\bf D}_n$ are less constrained since $\spr{\vecb}$ itself
satisfies $\spr{\vecb}^2(\eta)=k^2(\eta) \leq 1$ (equation
(\dr{c2})).

Let us consider the angular momentum of a circular loop of
invariant length $2\pi$ and hence corresponding total conserved
energy $E = 2 \pi m^2$.  Such a loop is given by
\be
\veca(q) = (\cos q , \sin q, 0)\; ; \qquad \vecb(\eta)=(k\cos
\eta, -k \sin \eta, 0),
\dle{circ}
\ee
where $k$ must be constant.  The loop oscillates between the
maximum and minimum radii of $(1 \pm k)/2$, so that for $k=0$ it
is stationary with fixed length $\pi$ (see the discussion above).
An energy $E=\pi m^2$ is stored in the string tension when $k=0$,
so the rest of
the energy must be stored in angular momentum ${\bf J}$:
\be
{\bf J} = m^2 \oint d\sigma (\vecx \wedge \dot{\vecx})
\ee
which is conserved by the equation of motion (\dr{eqnlecon}). On
substitution of $\veca$ and $\vecb$ from (\dr{circ}) this gives
\ba
J & = & \frac{C^2}{2\pi} \qquad (= NZ)
\nn
\\
& = & \frac{m^2}{4}(1-k^2) \nn
\ea
which is maximal for $k=0$ (vorton solution), and vanishes when
$k=1$ (NG limit).  As for a point particle moving in a circular
orbit, one can construct an effective potential for the loop
motion \dc{CPG}.  This has a contribution from the inward tension
$m^2$ and another from the centrifugal force.  Let $r(t)$ be the
radius of the loop at time $t$ so that $0\leq r \leq 1$. Then the
effective potential $\Upsilon(r,k)$ is given by \dc{CPG}
\ba
\Upsilon(r,k) & = & 2 \pi r m^2 + \frac{J}{r}
\nn
\\
& = & m^2 \left[ 2 \pi r + \frac{\pi}{2r}(1-k^2) \right]
\nn
\ea
which is plotted in figure \dr{pot} for different values of $k$.
Note that $\Upsilon = 2 \pi m^2$ at $r = (1 \pm k)/2$ as observed
above. In this chiral case, the situation is much more simple than
that studied in \dc{CPG} for strings with time- and space-like
currents:  here the loop motion is characterized by two parameters
$C$ and $E$ rather than three.

\begin{figure}[ht]
\begin{picture}(0,0)
\put(380,-180){\makebox(0,0)[lb]{ $r$}}
\put(380,-60){\makebox(0,0)[lb]{$\frac{\Upsilon(r,k)}{m^2}$}}
\put(130,-40){\makebox(0,0)[lb]{$k=0$}}
\put(148,-160){\makebox(0,0)[lb]{$k=1$}}
\put(115,-80){\makebox(0,0)[lb]{$k=$}}
\put(120,-90){\makebox(0,0)[lb]{$0.6$}}
\put(120,-130){\makebox(0,0)[lb]{$k=0.9$}}
\end{picture}
\centerline{\hbox{
\psfig{figure=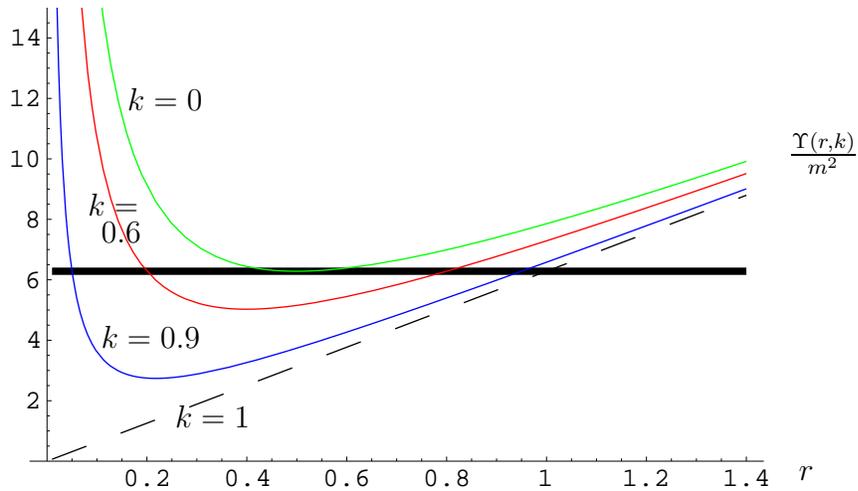}}}
\caption{Effective potential for loop motion as a function of $k$.
The thick solid line indicates the total conserved energy of the
system, measured in units of $m^2$.  The maximum radius is $r=1$.}
\label{pot}
\end{figure}

As we have noted, in general $k(\eta)$ need not be constant.  An
example of a loop solution for which this is the case is given by
\be
\veca(q) = (\cos q , \sin q, 0)\; ; \qquad \vecb(\eta)=(0,
-\frac{1}{2} \sin \eta, 0) \longleftrightarrow k^2(\eta)
=\frac{1}{4} \cos^2\eta,
\dle{simple}
\ee
see figure \dr{ksimple}.  This loop has angular momentum $J = m^2
\pi / 2 ( < C^2 / 2\pi)$ and does not self-intersect.  The figure
also indicates one of the two points on the loop for which $k=0$
(and so $|\dot{\vecx}| = |\vecx'|=1/2$): this point executes a
circular trajectory of radius $1/2$. Below we will see that any
loop with this form of $\vecb(\eta)$ does not self-intersect.

\begin{figure}[ht]
\begin{picture}(0,0)
\put(226,-43){\makebox(0,0)[lb]{$\bullet$}}
\put(274,-64){\makebox(0,0)[lb]{$\bullet$}}
\put(296,-112){\makebox(0,0)[lb]{$\bullet$}}
\put(274,-162){\makebox(0,0)[lb]{$\bullet$}}
\put(226,-183){\makebox(0,0)[lb]{$\bullet$}}
\end{picture}
\centerline{\hbox{
\psfig{figure=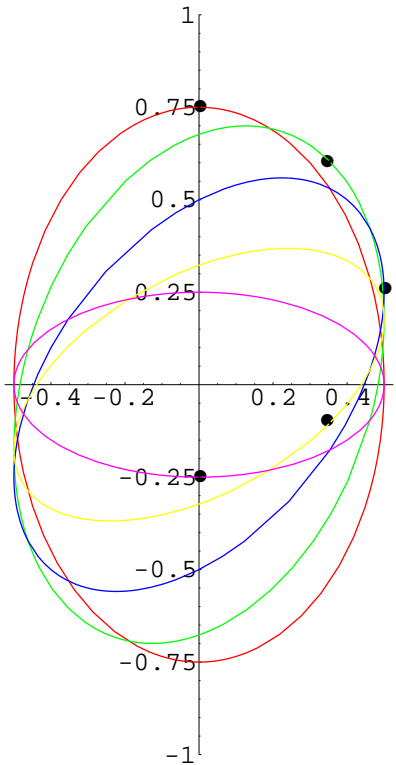}}}
\caption{Evolution of the loop given in (\dr{simple}) through half a period.
At $t=0$ the loop is symmetric about the vertical axis; at $t=L/2 = \pi$ it
is symmetric about the horizontal axis.  Intermediate times increase in
steps $\pi/8$.  At each time, a point on the loop
is labeled by a $\bullet$.  This is the point $k=0$ (there are 2 points
with $k=0$.  The other is not labeled and is diametrically opposite)
and it executes a circle of radius $1/2$.}
\label{ksimple}
\end{figure}

We now study the self-intersection probability of loops with
non-constant $k$.

\section{Self-intersection properties}\dle{s:kvar}

The self-intersection probability, $P_{int}$, of loops with given
numbers of harmonics on $\veca$ and $\vecb$ but constant $k$ was
studied in \dc{US}.  This was done through a simple adaptation of
the code of Siemens and Kibble \dc{SK} who studied the same
question for NG loops (i.e.\ when $k=1$).  Their work was in turn based on
methods developed by DeLanley et al \dc{DES,BD,BCD} who showed how, for a
fixed number of harmonics, the Fourier series (\dr{fourier}) could
be generated such that constraint (\dr{c1}) is satisfied.
Here we use a modified form of the same code
to study $P_{int}$ when $k(\eta)$ is not constant.

As seen in section \dr{s:eqn}, $k(\eta)$ can be any periodic
function provided $0 \leq k^2(\eta) \leq 1$.  Non-constant $k$ means
that the charge per unit length varies along the string and this
seems physically reasonable, especially for strings whose length
is larger than the horizon or for loops formed as the result of
self-intersection of other strings: fluctuations in charge will
occur during the phase transition which produces the strings, and
charge can be built up in self-intersections.

For non-constant $k(\eta)$, the self-intersection probability
$P_{int}$ might be expected to depend
on the number of zeros $n_{0}$ in the function $k(\eta)$ (since
when $k=0$ the loop never self-intersects), and also on
maximum amplitude, $A$, of $k(\eta)$.  The dependence of $P_{int}$ on
these parameters will be studied.

Unfortunately, once $k\neq$ constant, the freedom in possible loop
solutions increases since there is no longer any constraint on
the coefficients ${\bf C}_n$ and ${\bf D}_n$ in the Fourier
expansion of $\spr{\vecb}$ other than $0 \leq \spr{\vecb}^2 \leq
1$. One way to proceed is just to pick out by hand specific
functional forms of $k(\eta)$ (of which a constant is just one
case) and then try to construct all possible coefficients ${\bf
C}_n$ and ${\bf D}_n$ consistent with that $k(\eta)$, as was done
in \dc{DES,BD,BCD} for constant $k$\footnote{However, we have so
far been unable to generalize the methods of \dc{DES} to this
case. Any simple attempt always generated unwanted centre of mass
(constant) terms in the Fourier expansion (\dr{fourier}) of
$\spr{\vecb}$. For example, suppose one had generated a vector
$\spr{\bf d}$ of modulus 1 using \dc{DES}, and then set
$\spr{\vecb} = A \cos \eta \spr{\bf d}$ $(A<1)$.  This gives
$k^2(\eta) = A^2 \cos^2 \eta$. The problem is that the Fourier
expansion of $\spr{\vecb}$ now has a complicated constant term:
for example the term ${\bf C}_1 \cos \eta$ in the expansion of
$\spr{\bf d}$ leads to a constant term ${\bf C}_1 A / 2$ in the
Fourier expansion of $\spr{\vecb}$. Below we use a more simple
approach.}. One such simple function is
\be
k^2(\eta) = A^2 \cos^2 n \eta \qquad (A < 1)
\dle{simplek}
\ee
which has $2n$ zeros.  An example of a non-self-intersecting loop
with $n=1$ and $A=1/2$ is shown in figure \dr{ksimple}.  To see if
intersection is possible for any $A$ and $n$ recall that
self-intersection occurs if there is a solution to
\be
\veca( T+\sigma_1)+\vecb(T-\sigma_1) =
 \veca(T+\sigma_2)+ \vecb(T-\sigma_2)
\dle{sisec}
\ee
for some $0< \sigma_1 \neq \sigma_2 < L$ and $0 < T < L/2$. Let
$\chi$ and $\Phi$ be arbitrary angles and consider
\ba
\veca(q) \= \frac{1}{m}(\cos m q, \cos \chi \sin m q,\sin \chi
\sin m q),
\nn
\\
\vecb(\eta)\=\frac{A}{n}(\cos \Phi \sin n \eta, \sin \Phi \sin
n\eta, 0) ,
\dle{trial}
\ea
which gives $k(\eta)$ as in (\dr{simplek}). Now let $c=(\si_1 +
\si_2)/2$, $\delta = (\si_1 - \si_2)/2$, $q = T + c$ and $\eta =
T-c$.  Then the self intersection condition (\dr{sisec}) becomes
$$
\veca(q+\delta ) - \veca(q-\delta) = \vecb(\eta+\delta)-
\vecb(\eta-\delta)
$$
for which we must find solutions for $\eta,q,\delta$ with
$0<\delta < 2\pi$. On substitution of (\dr{trial}), this condition
becomes
\ba
\lefteqn{ \frac{1}{m}( -\sin mq \sin m \delta,\cos mq \sin m \delta \cos
\chi,\cos mq \sin m \delta \sin \chi )}
\nn
\\
& =& \frac{A}{n}(\cos \Phi
\cos n \delta \sin n \delta ,\sin \Phi \cos n\eta \sin n\delta,0)
\nn
\ea
for which the only solution is $\delta=0$.  Thus for $\vecb$ given
in (\dr{trial}) there are no self-intersections.

Let us instead consider a slightly more general form of
$\vecb(\eta)$;
\be
\vecb(\eta) = \frac{A}{n} (\sin n \eta, - \frac{\cos 2 d n \eta}{2
d} \cos \Phi , - \frac{\cos 2 d n \eta}{2
d} \sin \Phi )
\dle{bgen}
\ee
where $d$ is an integer greater than or equal to 1.  The
corresponding function $k^2(\eta)$ once again $2n$ zeros, but the
larger $d$ the more oscillations there are in $k^2(\eta)$ (figure
\dr{Kplot}).

\begin{figure}[ht]
\begin{picture}(0,0)
\put(230,-210){\makebox(0,0)[lb]{$\eta$}}
\put(100,-104){\makebox(0,0)[lb]{$k^2(\eta)$}}
\end{picture}
\centerline{\hbox{
\psfig{figure=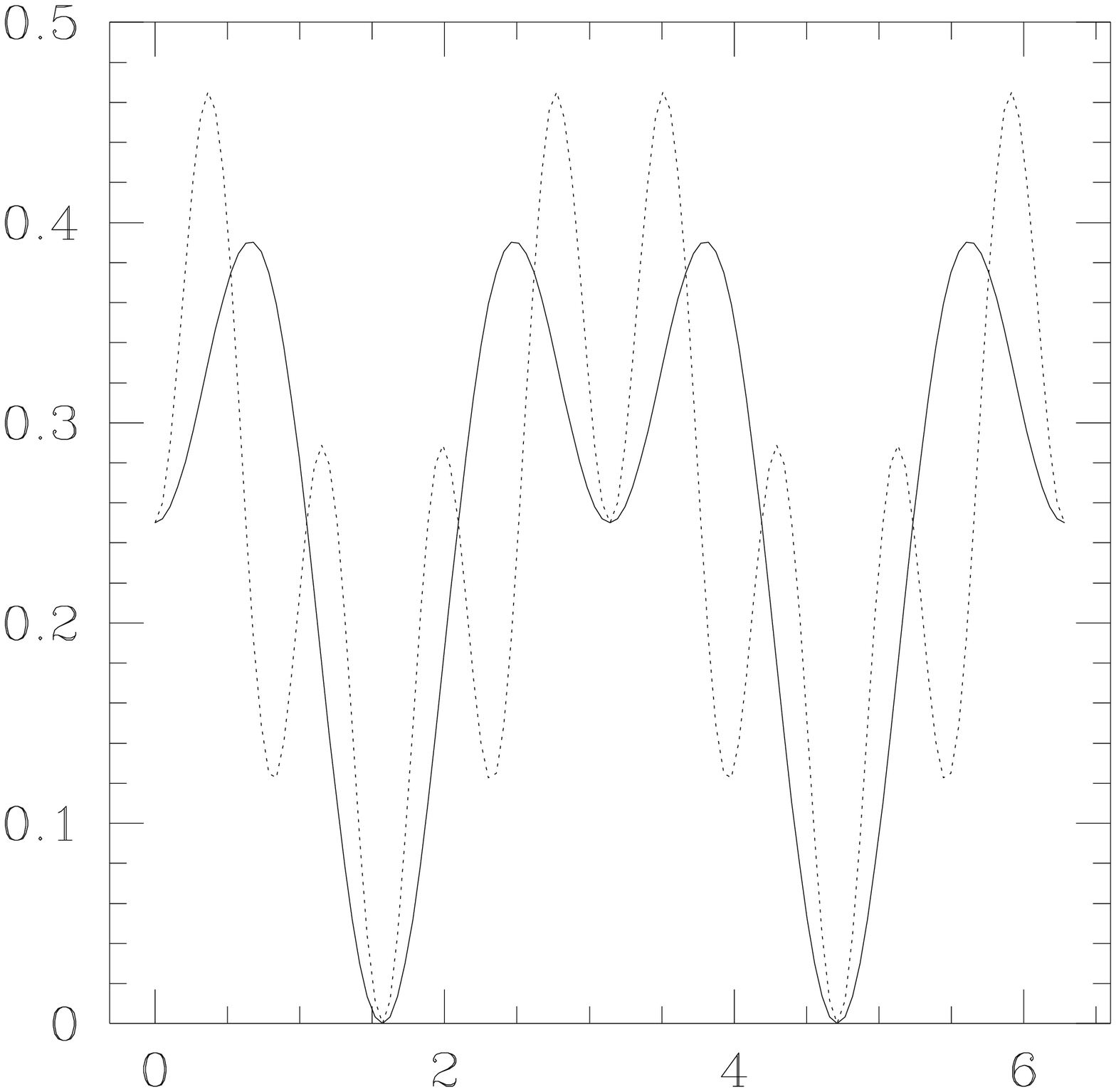,height=7.5cm,width=7.5cm}}}
\caption{Plot of $k^2(\eta) = A^2(\cos^2n\eta + \sin^2 2 d n \eta)$ for
$A=1/2$, $n=1$ and $d=1$ (solid line), $d=2$ (dotted line).}
\label{Kplot}
\end{figure}

The self-intersection condition now becomes (we set $\Phi=0$ for
simplicity)
\ba
\lefteqn{\frac{1}{m}( -\sin mq \sin m \delta,\cos mq \sin m \delta
\cos \chi,\cos mq \sin m \delta \sin \chi )}
\nn
\\
\= \frac{A}{n}(\cos n \eta \sin n \delta, \frac{1}{2d} \sin 2 d n
\eta \sin 2 d n \delta ,0)
\nn
\ea
which implies that
$$
\cos m q = 0 = \sin 2 d n \eta \qquad \Longleftrightarrow \qquad
\sin m q = \pm 1 = \cos 2 d n \eta \qquad (\longleftrightarrow
 \cos n \eta = 0,\pm 1),
$$
where $\delta$ must satisfy (for $\cos n \eta \neq 0$)
$$
\pm \frac{1}{m} \sin m \delta = \frac{A}{n} \sin n \delta.
$$
If $n$ and $m$ have no common factors there are solutions and
hence self-intersections.

\subsection{Numerical results}\dle{s:num}

The self-intersection probability, $P_{int}$, of loops with
$\vecb$ of the form given in (\dr{bgen}) was studied numerically.
For such loops $P_{int}$ is therefore a function of $n_0 = 2n$,
$d$, $A$, and also of $N_a$, the maximum number of harmonics on
the vector $\veca$. (This vector was generated using the methods
of \dc{DES}).
Note that in this case the charge $C$ is given by
\be
C = \frac{m}{2} \oint d\sigma \left[ 1 - A^2(\cos^2 n \sigma +
\sin^2 2 d n \sigma) \right]^{1/2},
\ee
which is essentially independent of the values of $n$ and $d$ for
$A \lsim 0.5$. Thus for a
given charge $C$ on the loop, the dependence of $P_{int}$ on $n$
and $d$ can be investigated, and also compared with the case in
which $k$ is constant \dc{US}.

Figure \dr{figNa} shows the dependence of $P_{int}$ on $N_a$ for
$n=2$ and $d=1$. Each point shown was obtained by generating 10
samples, each containing 100 loops, and looking for
self-intersections of each of these loops: the point is the
average number of self-intersections, and the error bar is the
standard deviation of this mean.  This is exactly the procedure
used by Siemens and Kibble, and details can be found in their
paper. For fixed $\vecb$ (hence fixed charge), the
self-intersection probability increases as the number of harmonics
in $\veca$ increases.  This is the expected behaviour as the loops
are more contorted for larger $N_a$. Interestingly $P_{int}$ is
only fractionally smaller here than that obtained in \dc{US} for
the same charge and constant $k$ (corresponding to $n_0 = 0$).

\begin{figure}[ht]
\centerline{\hbox{
\psfig{figure=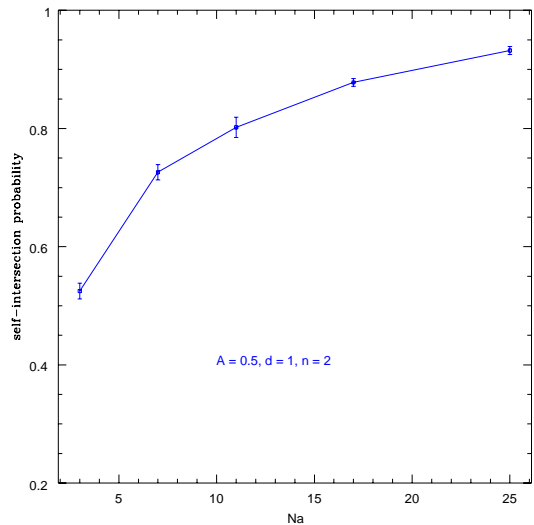,height=7.5cm,width=7.5cm}}}
\caption{Self-intersection probability as a function of $N_a$, the
number of harmonics on $\veca$.  All loops considered have the
same charge $C$.} \label{figNa}
\end{figure}

The left-hand plot in figure \dr{ndep} shows instead the effect of
fixing $N_a$ (=3) but increasing the number of zeros in $k(\eta)$.
The probability $P_{int}$ decreases as expected since each point
for which $k=0$ has a very constrained motion.  The graph shows
results for three different values of $A$ (or equivalently $C$):
as $C$ increases, $P_{int}$ decreases --- for a given $C$, the
self-intersection probability of a loop depends on the form of
$k(\eta)$. The right-hand plot of figure \dr{ndep} is similar to
the left-hand plot, and shows how, for fixed $C$, $P_{int}$
increases with $N_a$ but decreases with $n$.  These effects are
equally strong, in that if $N_a$=$n$, $P_{int}$ tends to a
constant value.  For comparison the results obtained in \dc{US}
for constant $k$ are plotted also.

Finally, we investigated the dependence of $P_{int}$ on $d$.
Figure \dr{ddep} shows that for fixed $n$, $A$ and $N_a$, the
self-intersection probability initially decreases as $d$ increases
but then seems to have an upturn. We are unable to explain this
behaviour at present.

\begin{figure}[ht]
\begin{picture}(0,0)
\put(255,-28){\makebox(0,0)[lb]{${\bf{\circ}}$}}
\put(255,-50){\makebox(0,0)[lb]{$\circ$}}
\end{picture}
\centerline{\hbox{
\psfig{figure=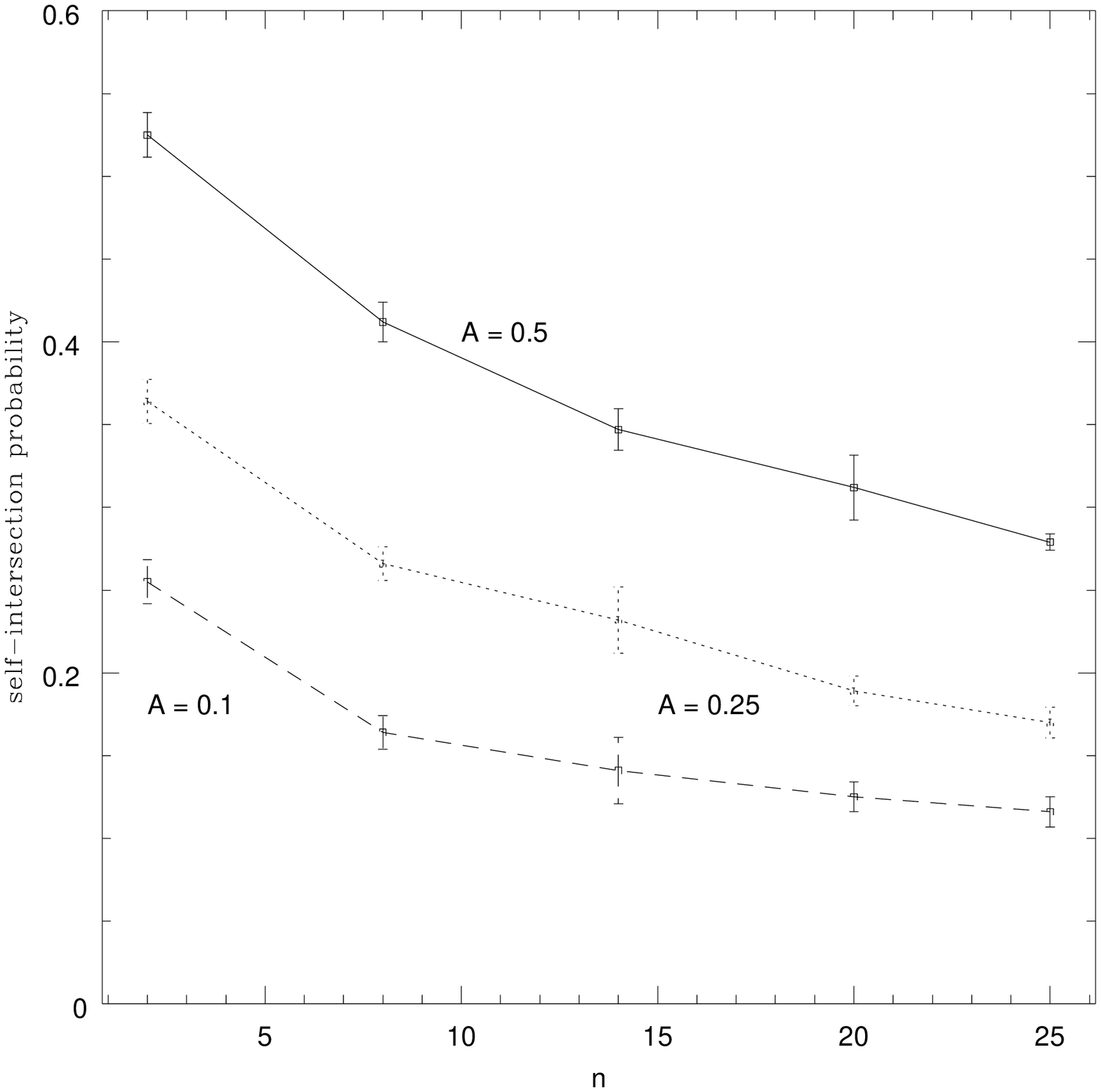,height=7.5cm,width=7.5cm}
\psfig{figure=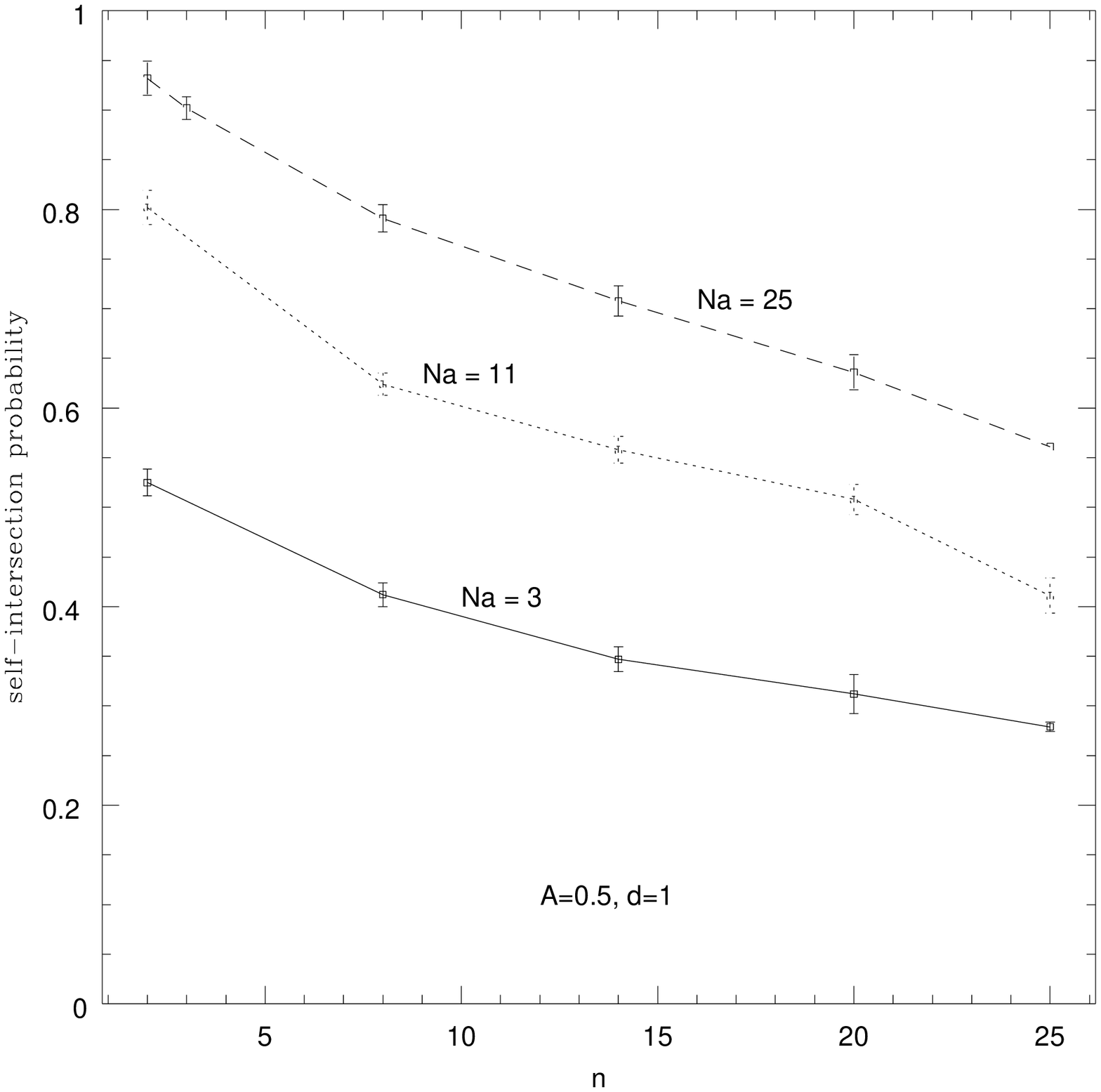,height=7.5cm,width=7.5cm}}} \caption{a)
This figure shows how, for a given charge $C$ (determined by $A$),
the self-intersection probability decreases with $n$.  b) The
dependence on $N_a$ and $n$ for fixed $C$. For comparison we have
also plotted the results obtained in \dc{US} for constant $k$
(upper circle: $N_a$=25, lower circle $N_a$=11). } \label{ndep}
\end{figure}

\begin{figure}[ht]
\centerline{\hbox{
\psfig{figure=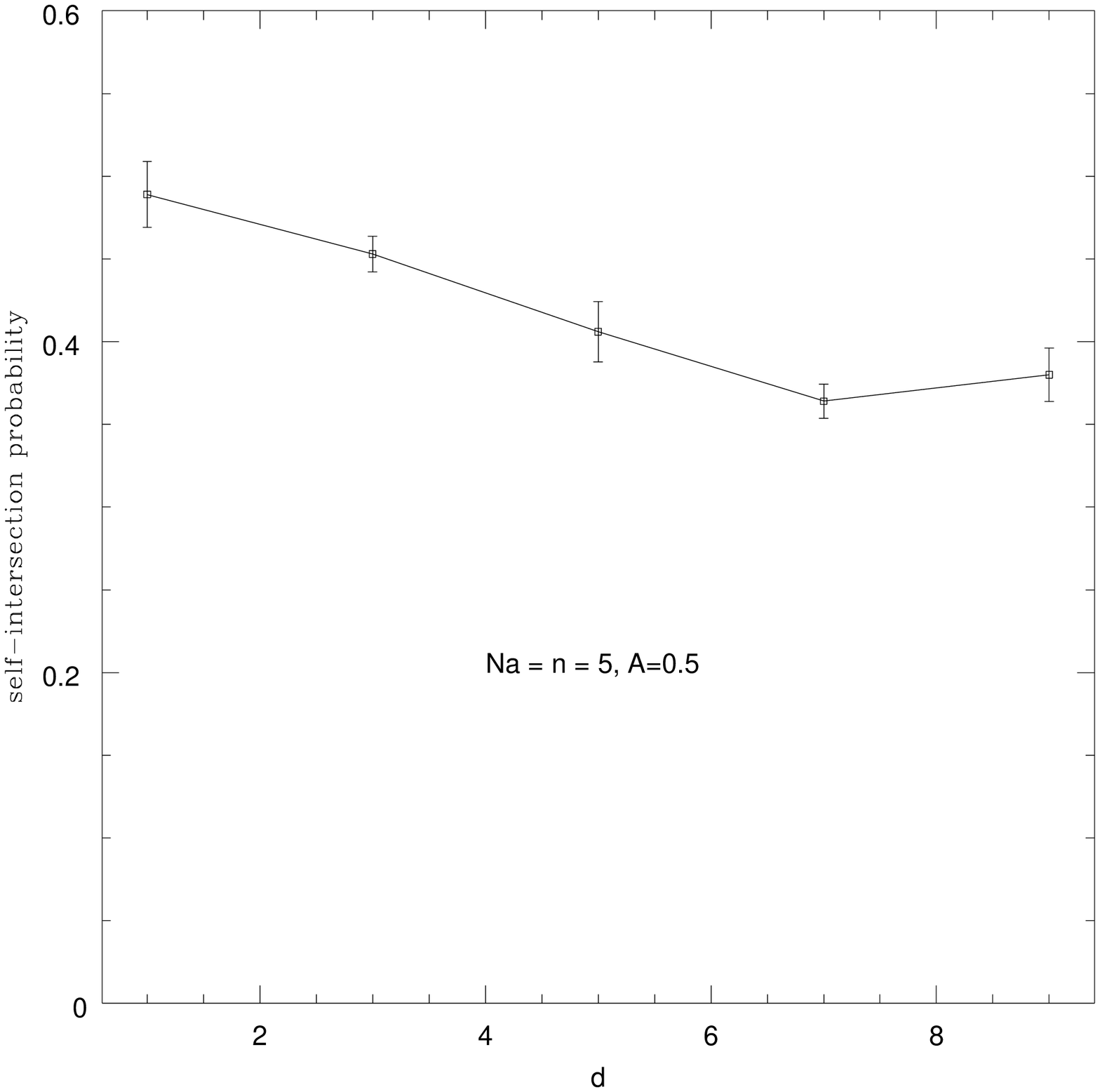,height=7.5cm,width=7.5cm}}}
\caption{Dependence of $P_{int}$ on $d$.}
\label{ddep}
\end{figure}

\section{Conclusions}\dle{s:conc}

In this paper we have attempted study and clarify a number of
points regarding the evolution and gravitational properties of chiral
cosmic strings.  As was summarized in section \dr{s:2}, the
crucial difference between
the equations of motion for NG and chiral cosmic strings is the
constraint on the vector $\spr{\vecb}$: for NG strings
$\spr{\vecb}^2(\eta) = 1$ $\forall \eta$, whereas for chiral strings
$\spr{\vecb}^2(\eta) (=k^2(\eta)) \leq 1$.  Equation (\dr{Cr}) shows that
$k^{2}(\eta)$ determines the charge on the chiral string.

We saw in section \dr{s:gen} that chiral strings with $k=0$ ($\forall
\eta$) move along themselves and never self-intersect.  If the string
forms a loop, the energy of this arbitrary shaped vorton is
equipartitioned between tension and angular momentum.  The charge on
the vortons is given by $C = m L_{{\rm phys}}$ where
$ L_{{\rm phys}}$ is the constant physical length of the vorton.

Infinite straight chiral strings were studied in section \dr{s:metric}.  We
saw that the energy momentum tensor contains non-diagonal terms
$T^{tz} \neq 0$.  These represent the momentum along the string.
Furthermore, $T^{tt} \neq T^{zz}$ (if $k \neq 1$) which is reminiscent
of the situation which occurs with wiggly NG strings.  As a
consequence of the form of $T^{\mu \nu}$, the weak-field metric around
the string was shown to contain a $dt \, dz$ term which means that
photons (and relativistic particles) moving near the string are
dragged in the direction of the string.  We also observed that there
is a $k$-dependent deficit angle as well as a $k$-dependent Newtonian
potential.

Regarding the evolution of a chiral cosmic string network (which could
formed at the end of D-term inflation), it is important to understand
whether or not the loops can self-intersect and then decay.  If they
cannot decay, this would lead to a cosmological catastrophe as they
would dominate the energy density of the universe.  In section
\dr{ss:ang} we studied the effective potential for the motion of a non
self-intersecting circular loop for which $0 \leq k < 1$.  In section
\dr{s:kvar} we considered loops with non-constant $k$: the physical
reason for which one might expect $k$ not to be constant is that
charge will build up as a result of self-intersections, and also
fluctuate during the phase transition which forms the string.
Analysis of specific form of $k(\eta)$ (given via (\dr{bgen})) showed
that self-intersection is possible for these loops.  The ensuing
numerical analysis showed that the self-intersection probability
depends on the form of $k(\eta)$ and is not uniquely determined by the
charge $C$ of the loop.
This unfortunately suggests that
even if one were able to estimate $C$
for the strings in a chiral cosmic string network, this would not be
sufficient to determine the self-intersection properties of the loops.
As a further
problem it still remains to understand the fate of the daughter loops.

A number of interesting questions remain to be studied.  Regarding the
metric (section \dr{s:metric}), it would be interesting to go beyond the
weak-field approximation and also to study carefully the potential
cosmological consequences of the $dt \, dz$ term \dc{StVa}.  This cross-term is
the main difference between the metric for NG and chiral
strings.  Concerning the evolution of a network of chiral cosmic
strings it is clear that if the network is formed with $k(\eta) = 0$
$\forall \eta$ and for all strings, then this leads to a cosmological
catastrophe: this is the only case in which the answer for $P_{int}$
is unique and zero! --- the strings cannot self-intersect and are
frozen.  Similar problems occur if this state is reached at anytime
during the evolution of the network.  This vorton problem was studied in
\dc{CD} where it was noted that the
quantum number $C$ should be larger for chiral strings than for
strings with time- or space-like currents.
However, work still needs to be done to see if $C$ is maximal or not
\dc{DP}.  If it is not maximal (i.e.\ $k \neq 0$ $\forall \eta$) it
still remains to understand the ultimate fate of the daughter loops, and
hence that of the network itself.

\section*{Acknowledgements}

I am particularly grateful to Tanmay Vachaspati for critically reading
a previous version of this paper, for interesting correspondence, and
for much advice and encouragement. I would also like to thank Patrick
Peter who spotted a mistake in $T^{\mu
\nu}$, again in a previous version of this paper; and Tom
Kibble for useful comments on that version. My thanks also to Ola
T\"ornkvist for a useful discussion, and finally I must
mention Mike Pickles and Anne Davis who have since told me that
they may be able to generalize the methods of \dc{BD} to
non-constant $k$'s. This work was supported in part by the Swiss
NSF and an ESF network.

\typeout{--- No new page for bibliography ---}

\end{document}